\documentclass[prl,twocolumn,aps,floats,superscriptaddress,
floatfix,showpacs,showkeys,amsmath,amssymb]{revtex4}

\usepackage{dcolumn}
\usepackage{bm}
\usepackage{graphicx}
\usepackage{times}

\begin{document}
\title{The analogue Hawking effect in rotating polygonal 
hydraulic jumps} 
\author{Arnab K. Ray}\email{arnab.kumar@juet.ac.in}
\affiliation{Department of Physics, Jaypee University
of Engineering \& Technology, 
Raghogarh, Guna 473226, Madhya Pradesh, India}
\author{Niladri Sarkar}\email{nsarkar@pks.mpg.de}
\affiliation{Max-Planck Institut f\"ur Physik Komplexer 
Systeme, N\"othnitzer Str. 38, Dresden, D-01187 Germany}
\author{Abhik Basu}\email{abhik.basu@saha.ac.in}
\affiliation{Condensed Matter Physics Division, Saha
Institute of Nuclear Physics, Calcutta 700064, India}
\author{Jayanta K. Bhattacharjee}\email{jkb@hri.res.in}
\affiliation{Harish Chandra Research Institute, Chhatnag 
Road, Jhunsi, Allahabad 211019, India}
\date{\today}
\begin{abstract}
Rotation of non-circular hydraulic jumps is a recent
experimental observation that lacks a theory based on 
first principles. Here we furnish a basic theory of this 
phenomenon founded on the shallow-water model of the circular 
hydraulic jump. The breaking of the axial symmetry morphs 
the circular jump into a polygonal state. Variations 
on this state rotate the polygon in the azimuthal direction. 
The dependence of the rotational frequency on the flow rate 
and on the number of polygon vertices agrees with known 
experimental results. We also predict how the rotational 
frequency varies with viscosity. Finally, we establish 
a correspondence between the rotating polygonal structure 
and the Hawking effect in an analogue white hole. The 
rotational frequency of the polygons affords a direct 
estimate of the frequency of the thermal Hawking radiation. 
\end{abstract}


\maketitle
In low-dimensional flows, the hydraulic jump is 
associated with an abrupt increase in the height of a
flowing liquid~\cite{ll87}.  
Along a standing circular locus, this feature 
is known as the two-dimensional circular 
hydraulic jump~\cite{bdp93,rb07}. Its
geometry maintains a symmetry about a reference axis that 
passes normally through the point of the origin of the flow 
on a flat plane. To develop a cogent mathematical theory, 
the flow is viewed as a shallow layer of liquid~\cite{bdp93}  
confined to a horizontal plane, diverging radially 
outwards from a source point. Thereafter, 
applying the boundary-layer approximation in the shallow
flow, viscosity is invoked~\cite{watson,bdp93}, along with
all other complications of the nonlinear Navier-Stokes 
equation. 
A compelling evidence in favour of viscosity comes from 
an experimentally verified formula of the jump 
radius, scaled in terms of viscosity~\cite{bdp93,hansen97}. 
Various physical interpretations of this scaling law 
have been forwarded~\cite{bdp93,rb07}.

Circular hydraulic jumps appear in Type-I and Type-II 
states~\cite{ehhhmbhw2}, of which the latter is formed by 
increasing the height of the flow in the post-jump region. 
Type-II jumps have a wider jump region with a surface eddy 
all along its circular 
rim like a floating torus~\cite{ehhhmbhw2}. 
When water is replaced with a liquid of much 
greater viscosity, the Type-II state 
spontaneously breaks the axial symmetry of the 
circular state, resulting in a front that has a polygonal
geometry~\cite{behh96,ehhhmbhw1,ehhhmbhw2,mwb12}. 
This transition has visibly temporal features 
because a linear instability arises from 
the initially stable circular
state, and leads to the formation of a non-axisymmetric
polygonal structure, whose dependence on the azimuthal
angle is periodic~\cite{mwb12}. 
Within the widened jump region, the
transport of liquid also becomes azimuthal, although 
prior to it, in the pre-jump region, the flow 
is radial~\cite{mwb12}. Clearly, the 
breaking of axial symmetry is localized only at the 
jump radius or thereabouts. All of these features 
of the polygonal jump have been known well for 
two decades, but in addition, as reported very recently,  
the polygonal jump undergoes a rotation~\cite{tm15}.
This new phenomenon has been named the ``rotational hydraulic 
jump"~\cite{tm15}~\footnote{A hexagonal feature around 
Saturn's North Pole rotates~\cite{god88}.}. 

Thus far we have provided a brief narrative of the role 
that viscosity plays in the multifarious properties of 
the two-dimensional hydraulic jump. Now we consider 
the jump from a different perspective~\cite{su02,us05,
vol05,sbr05,rb07}, according to which the position of 
the standing jump is a boundary where the steady radial 
velocity, $v_0(r)$, equals the local speed of 
long-wavelength surface 
gravity waves, $\sqrt{gh_0}$, with $h_0(r)$ being the 
steady local height of the flow layer. This boundary 
is like a standing horizon,
segregating the supercritical and the subcritical regions
of the flow, with criticality referring to the condition
under which the speed of the bulk flow matches the speed
of surface gravity waves~\cite{rb07}. This entire 
point of view lays strong emphasis on the advective
and pressure terms in the Navier-Stokes equation, to 
the neglect of the viscous term. The radius of the 
horizon (which is also the jump), $r_{\mathrm J}$, is 
defined by the critical condition, 
$v_0^2(r_{\mathrm J})=gh_0(r_{\mathrm J})$, and 
sets a spatial limit for the transmission of information. 
As the equatorial flow proceeds from its point of origin, 
its radial velocity, $v_0 > \sqrt{gh_0}$, but viscosity
and the radial geometry slow down the flow. When the 
critical condition is met, both the jump and the horizon 
occur simultaneously~\cite{rb07}. 
In the supercritical part 
of the jump, where $v_0 > \sqrt{gh_0}$, gravity waves 
(as carriers of information) 
cannot travel upstream against the bulk flow, and hence
every point in the supercritical region remains uninformed 
about the fate of the flow downstream. This state of 
affairs prevails everywhere within the jump, and so it
acts as an impenetrable barrier against the percolation 
of any information from the outside to the inside. 
The horizon implied by the critical 
condition has been amply demonstrated~\cite{jpmmr}, 
but the horizon by itself is inadequate to explain why 
a jump should coincide with it, a point that has been 
qualitatively addressed and appreciated for long~\cite{ll87}.   
Concisely stated, the horizon is a necessary condition 
for the jump but not a sufficient one, and so without 
reference to anything regarding the jump, the horizon 
is just an analogue of a white hole. In this study 
we show that the formation of a polygonal jump 
and its observed rotation~\cite{tm15} are the natural 
outcomes of breaking the axial symmetry of the circular 
horizon of the analogue white hole. The breaking of the 
axial symmetry, localized near the horizon,
is due to the analogue surface gravity. 
The rotation has a persuasive similarity with 
the Hawking effect, which is also due to the analogue 
surface gravity. 

The mathematical description of the two-dimensional flow 
is most succinctly framed in the cylindrical coordinate 
system, $(r,\phi,z)$, and by tailoring the Navier-Stokes equation 
accordingly~\cite{ll87}. Our analysis starts with the steady 
height-integrated Navier-Stokes equation of a shallow-water 
radial base flow~\cite{bdp93}, 
\begin{equation}
\label{bdpns}
v_0\frac{{\mathrm d}v_0}{{\mathrm d}r}+ 
g\frac{{\mathrm d}h_0}{{\mathrm d}r}=-\nu \frac{v_0}{h_0^2}, 
\end{equation}
and the steady height-integrated equation of continuity, 
\begin{equation}
\label{bdpcon}
\frac{1}{r}\frac{\mathrm d}{{\mathrm d}r}
\left(rv_0h_0\right)=0, 
\end{equation} 
in its differential form. The integral form of Eq.~(\ref{bdpcon}) 
is $rv_0h_0 = Q/2\pi$, where $Q$, a constant, is the steady 
volumetric flow rate. The subscript ``$0$" in the foregoing 
equations stands for steady radially varying  
quantities, of which the velocity of the flow, $v_0(r)$, has
been obtained in the shallow-water theory by vertically averaging 
the radial component of the velocity across the height of the 
flow. The boundary conditions used for the averaging are that 
velocities vanish at $z=0$ (the no-slip condition), and vertical 
gradients of velocities vanish on the free surface of the flow 
(the no-stress condition)~\cite{bdp93,bpw97,sbr05}. These 
boundary conditions 
are applied under the standard assumption that while the vertical 
velocity is much small compared with the radial velocity, the 
vertical variation of the radial velocity (through the shallow layer 
of water) is much greater than its radial variation~\cite{bdp93}.
In this manner all dependence on the $z$-coordinate is 
averaged out. 

Now, about the steady radial background flow, as implied by 
Eqs.~(\ref{bdpns})~and~(\ref{bdpcon}), we develop a 
time-dependent azimuthal perturbation scheme 
prescribed by $h(t,r,\phi)=h_0(r)+
h^\prime(t,r,\phi)$, $v_r(t,r,\phi)=v_0(r)+v_r^\prime(t,r,\phi)$
and $v_\phi(t,r,\phi)=0+v_\phi^\prime(t,r,\phi)$. 
All primed quantities are time-dependent 
perturbations in both the radial and azimuthal coordinates, 
with $v_r^\prime$ and $v_\phi^\prime$ being perturbations on 
the radial and azimuthal velocity components, respectively. 
We have designed the azimuthal flow to be entirely a 
perturbative effect without any presence in the steady 
background flow. Our next step
is to define a new variable, $f=rhv_r$, whose steady value, 
$f_0=rh_0v_0$, is a constant, as we can see 
from Eq.~(\ref{bdpcon}).  
Within a multiplicative factor, owed to the two-dimensional
geometry of the system, $f_0$ gives the conserved volumetric
flow rate under steady conditions. So $f^\prime$ is a 
perturbation on this constant background volumetric 
flow rate. Linearizing in $f^\prime$ under the formula, 
$f=f_0+f^\prime$, gives us
$f^\prime = r\left(h_0v_r^\prime + v_0h^\prime \right)$. 
A similar linearization of the general 
time-dependent continuity equation, bearing both radial 
and azimuthal variations~\cite{ll87}, leads to 
\begin{equation} 
\label{hpertur} 
\frac{\partial h^\prime}{\partial t}=- \frac{1}{r} 
\frac{\partial f^\prime}{\partial r} - \frac{h_0}{r} 
\frac{\partial v_\phi}{\partial \phi}. 
\end{equation} 
Likewise, from the radial and azimuthal components of 
the Navier-Stokes equation, as expressed 
in cylindrical coordinates~\cite{ll87}, we derive  
\begin{equation}
\label{vrpertur1} 
\frac{\partial v_r^\prime}{\partial t} + 
\frac{\partial}{\partial r}\left(v_0 v_r^\prime\right) 
+g \frac{\partial h^\prime}{\partial r} =0 
\end{equation} 
and 
\begin{equation} 
\label{vphipertur1} 
\frac{\partial v_\phi^\prime}{\partial t} + 
v_0 \left(\frac{\partial v_\phi^\prime}{\partial r} + 
\frac{v_\phi^\prime}{r}\right) + \frac{g}{r} 
\frac{\partial h^\prime}{\partial \phi} =0, 
\end{equation}
respectively. We stress here that in extracting 
Eqs.~(\ref{vrpertur1})~and~(\ref{vphipertur1}), we have 
ignored all product terms of viscosity and the primed 
quantities. Therefore, in our treatment, viscosity makes 
its mark 
through the zero-order stationary quantities, $v_0$ and 
$h_0$, as Eqs.~(\ref{bdpns})~and~(\ref{bdpcon}) indicate. 
In all first-order equations involving the primed 
quantities, 
Eqs.~(\ref{hpertur}),~(\ref{vrpertur1})~and~(\ref{vphipertur1}),
viscosity does not appear explicitly, and 
exerts its influence implicitly through the zero-order
coefficients. This is a satisfactory approximation, to the 
extent that our principal concern is the effect of azimuthal
variations on the axially symmetric radial flow, and 
the breaking of the axial symmetry therefrom.

We use Eq.~(\ref{hpertur}) to substitute  
$\partial h^\prime/\partial t$ in the first-order partial 
time derivative of $f^\prime$, whereby 
we also obtain 
\begin{equation} 
\label{vrpertur} 
\frac{\partial v_r^\prime}{\partial t}=\frac{v_0}{f_0}
\frac{\partial f^\prime}{\partial t}+\frac{v_0^2}{f_0} 
\frac{\partial f^\prime}{\partial r}+\frac{h_0v_0^2}{f_0}
\frac{\partial v_\phi}{\partial \phi}.
\end{equation} 
Collectively, Eqs.~(\ref{hpertur}) 
and~(\ref{vrpertur}) present a closed set of conditions 
that express $h^\prime$ and $v_r^\prime$ as a linear 
combination of $f^\prime$ and $v_\phi$. At this stage 
we require two independent mathematical conditions on which 
we can impose Eqs.~(\ref{hpertur}) and~(\ref{vrpertur}). 
Such conditions are readily supplied by the perturbations 
of the radial and azimuthal dynamics, as shown in  
Eqs.~(\ref{vrpertur1})~and~(\ref{vphipertur1}), respectively. 
We take the second-order partial time 
derivative of these two coupled equations, and on them
we apply the conditions provided by 
Eqs.~(\ref{hpertur}) and~(\ref{vrpertur}), along with 
the second-order partial time derivative of Eq.~(\ref{vrpertur}). 
This long algebraic exercise ultimately delivers two 
equations that are second-order in time. They are
\begin{multline} 
\label{fpert2} 
\frac{\partial}{\partial t}
\left(v_0\frac{\partial f^\prime}{\partial t}\right)
+\frac{\partial}{\partial t}
\left(v_0^2\frac{\partial f^\prime}{\partial r}\right) 
+\frac{\partial}{\partial r}
\left(v_0^2\frac{\partial f^\prime}{\partial t}\right) \\
+\frac{\partial}{\partial r}\left[v_0
\left(v_0^2-gh_0\right)\frac{\partial f^\prime}{\partial r} 
\right] = -\frac{\partial}{\partial t}
\left(v_0^2 h_0\frac{\partial v_\phi}{\partial \phi}\right) \\ 
-\frac{\partial}{\partial r}\left[v_0h_0
\left(v_0^2-gh_0\right)\frac{\partial v_\phi}{\partial \phi} 
\right], 
\end{multline} 
derived from the perturbation of the radial dynamics, as given 
by Eq.~(\ref{vrpertur1}), and 
\begin{equation} 
\label{vazipert2} 
\frac{\partial^2 v_\phi}{\partial t^2}+\frac{v_0}{r}
\frac{\partial}{\partial r}\left(r\frac{\partial v_\phi}
{\partial t}\right)-\frac{gh_0}{r^2}
\frac{\partial^2 v_\phi}{\partial \phi^2} = 
\frac{g}{r^2}\frac{\partial}{\partial \phi}
\left(\frac{\partial f^\prime}{\partial r}\right), 
\end{equation} 
derived likewise from the perturbation of the azimuthal dynamics, 
as given by Eq.~(\ref{vphipertur1}). The two foregoing equations, 
linearized and coupled, form a set of wave equations in 
$f^\prime$ and $v_\phi$. A familiar wave equation in $f^\prime$ 
only, pertaining just to the radial dynamics, is obtained when 
all the $\phi$-derivatives on the right hand side
of Eq.~(\ref{fpert2}) are made to vanish~\cite{rb07}. Thereafter, 
the expression on the left hand side of Eq.~(\ref{fpert2})
is rendered compactly as
$\partial_\alpha \left( {\mathsf{f}}^{\alpha \beta}
\partial_\beta f^{\prime}\right) =0$, 
in which the Greek indices run from $0$ to $1$, with 
$0$ standing for $t$ and $1$ standing for $r$.
This simplification establishes an acoustic metric and an 
acoustic horizon in the physical problem of the two-dimensional 
hydraulic jump, the details of which have been discussed in a 
previous work~\cite{rb07}. Of course, we must remember that this
reasoning is valid only for an inviscid fluid in a potential 
flow. In our present approach, the steady background flow is
affected by viscosity, while the first-order perturbation has 
an azimuthal component, which we consider to be a major
advancement. It cannot be ignored at the jump front, 
as far as the formation of polygonal structures is concerned.

We anticipate solutions that are separable in $t$, $r$ and 
$\phi$, for the two linearly coupled wave equations, given 
by Eqs.~(\ref{fpert2})~and~(\ref{vazipert2}). In keeping 
with this stipulation, we set down 
$f^\prime (t,r,\phi)=A\exp[-i\omega t +i\Theta (r)+im\phi]$ 
and $v_\phi (t,r,\phi)= B\psi (r)\exp[-i\omega t +i\Theta (r)+ 
im\phi] \equiv B f^\prime \psi (r)/A$, in which $A$ and $B$ 
are constants. The factor of $e^{im \phi}$ captures the 
azimuthal variation. Along the radial direction, $f^\prime$
has a slow variation, except near the acoustic horizon. 
The horizon is significant because it is also the position 
of the hydraulic jump, where the axial symmetry 
of the circular jump front is broken, a polygonal structure 
emerges and the flow acquires an azimuthal component. 
Accordingly, $v_\phi$ is most conspicuous about the jump 
radius and decays sharply away from it, a feature that 
is captured by a strong local prominence for $\psi (r)$ at 
$r=r_{\mathrm J}$. This radial position also allows us to 
exploit the condition of the acoustic horizon, $v_0^2=gh_0$, 
greatly simplifying our analysis. And so Eqs.~(\ref{fpert2}) 
and~(\ref{vazipert2}) yield two characteristic equations as 
\begin{multline} 
\label{char1} 
\frac{B}{A}\psi (r_{\mathrm J})\left[m\omega v_0h_0 +imh_0 
\frac{\mathrm d}{{\mathrm d}r}\left(v_0^2-gh_0\right) \right]
=\omega^2 \\
+\omega\left[2i\frac{{\mathrm d}v_0}{{\mathrm d}r} -2v_0
\frac{{\mathrm d}\Theta}{{\mathrm d}r}\right] 
-i\frac{{\mathrm d}\Theta}{{\mathrm d}r} 
\frac{\mathrm d}{{\mathrm d}r}\left(v_0^2-gh_0\right),  
\end{multline} 
and
\begin{multline}  
\label{char2} 
\frac{B}{A}\psi (r_{\mathrm J})
\left[\omega^2 +
\omega\left(i\frac{v_0}{r_{\mathrm J}}+i\frac{v_0}{\psi}
\frac{{\mathrm d}\psi}{{\mathrm d}r}
-v_0\frac{{\mathrm d}\Theta}{{\mathrm d}r}\right) 
-\frac{gm^2h_0}{r_{\mathrm J}^2}\right] \\
= \frac{gm}{r_{\mathrm J}^2} 
\frac{{\mathrm d}\Theta}{{\mathrm d}r},
\end{multline} 
respectively. We emphasize that $v_0 (r_{\mathrm J})$, 
$h_0 (r_{\mathrm J})$ and their derivatives in 
Eqs.~(\ref{char1})~and~(\ref{char2}) carry their values 
only at $r_{\mathrm J}$. The consistency of $B\psi/A$ on 
the left hand sides of Eqs.~(\ref{char1})~and~(\ref{char2}), 
which are both quadratic in $\omega$, lets us eliminate 
$A/B$, resulting in a single quartic equation in $\omega$,
going as $(\omega^3 +\Gamma_2\omega^2 +\Gamma_1\omega 
+\Gamma_0)\omega =0$. One root of the quartic 
equation is $\omega =0$, which leaves behind a residual 
cubic equation. Marginal stability (signalling the onset 
of a possible instability) extracts another root of $\omega=0$ 
from the cubic equation, something that is possible only 
when $\Gamma_0 =0$, and which in turn leads to a critical 
value of $m=m_{\mathrm c}$. From the vanishing of the real 
part of $\Gamma_0$ we get 
\begin{equation} 
\label{mcrit1} 
m_{\mathrm c}^2 
= -\frac{r_{\mathrm J}}{gh_0}
\frac{\mathrm d}{{\mathrm d}r}\left(v_0^2 -gh_0\right) 
\left(1+ \frac{r_{\mathrm J}}{\psi}
\frac{{\mathrm d}\psi}{{\mathrm d}r}\right),  
\end{equation} 
and similarly the vanishing of the imaginary part of 
$\Gamma_0$ gives 
\begin{equation} 
\label{mcrit2} 
m_{\mathrm c}^2 = 
\frac{v_0 r_{\mathrm J}^2}{2gh_0} 
\left(\frac{{\mathrm d}\Theta}{{\mathrm d}r}\right)^2 
\left(\frac{{\mathrm d}v_0}{{\mathrm d}r}\right)^{-1}
\frac{\mathrm d}{{\mathrm d}r}\left(v_0^2 -gh_0\right),
\end{equation} 
which determines  
${\mathrm d}\Theta/{\mathrm d}r$ at the horizon. 

Having obtained $m_{\mathrm c}$ from the condition of 
marginal stability, we now cause a slight perturbation, 
$m=m_{\mathrm c}+\Delta m$ (with $\Delta m/m_{\mathrm c} \ll 1$), 
to effect a very small change in $\omega$ from $\omega =0$. 
Smallness of the value of $\omega$ allows us to ignore 
its higher orders in the cubic equation and retain only 
$\Gamma_1 \omega + \Gamma_0 \simeq 0$. Hence, 
$- \omega \simeq \Gamma_0/\Gamma_1$. On linearizing 
in $\Delta m$, we get 
\begin{equation}
\label{gamma0} 
\Gamma_0 = \frac{2m_{\mathrm c}gh_0v_0}{r_{\mathrm J}^2}
\left(\frac{{\mathrm d}\Theta}{{\mathrm d}r}-i\frac{2}{v_0}
\frac{{\mathrm d}v_0}{{\mathrm d}r}\right) \Delta m, 
\end{equation} 
while we write $\Gamma_1 = P+iQ$, in which, aided by some 
simplifications due to Eqs.~(\ref{mcrit1})~and~(\ref{mcrit2}),  
\begin{align*} 
P&= 
v_0^2\left(\frac{{\mathrm d}\Theta}{{\mathrm d}r}\right)^2
\left[2+\frac{{\mathrm d}\left(gh_0\right)/{\mathrm d}r}
{{\mathrm d}\left(v_0^2\right)/{\mathrm d}r}\right],\\
Q&=-\frac{{\mathrm d}\Theta}{{\mathrm d}r}\left[
\frac{\mathrm d}{{\mathrm d}r}\left(2v_0^2 -gh_0\right) 
+\frac{2v_0^2}{r_{\mathrm J}}
\left(1+ \frac{r_{\mathrm J}}{\psi}
\frac{{\mathrm d}\psi}{{\mathrm d}r}\right)\right].
\end{align*} 
For convenience we write $-\omega = \gamma \Delta m$, from 
which we get 
\begin{multline} 
\label{gammaexp} 
\gamma = \frac{2m_{\mathrm c}v_0gh_0}{r_{\mathrm J}^2\left(
P^2+Q^2\right)} \times \\
\left[\left(P\frac{{\mathrm d}\Theta}{{\mathrm d}r}
-\frac{2Q}{v_0}\frac{{\mathrm d}v_0}{{\mathrm d}r}\right)
-i\left(Q\frac{{\mathrm d}\Theta}{{\mathrm d}r}+
\frac{2P}{v_0}\frac{{\mathrm d}v_0}{{\mathrm d}r}\right)\right], 
\end{multline} 
whose form is best read as $\gamma = \Re(\gamma)+i\Im(\gamma)$. 
Now, the phase of the wave solution is 
$\exp[-i\omega t +i\Theta (r)+im\phi]$, from which we extract 
only the time-dependent part and recast it 
as $\exp[i\Re(\gamma) \Delta m\,t]
\times \exp[-\Im(\gamma) \Delta m\,t]$. The conclusion 
we draw is that $\Re(\gamma)$ causes the wave to travel along 
the $\phi$ coordinate (the azimuthal direction), and $\Im(\gamma)$, 
depending on its sign, causes either a growth or a decay in the 
amplitude of the travelling wave. To examine the latter feature, 
we look at Eqs.~(\ref{mcrit1})~and~(\ref{mcrit2}), both of which
give $m_c^2$ as a perfect square. As such, $m_c$ will have 
two roots of the same value but opposite signs. Since $\Im(\gamma)$
depends on $m_c$, as Eq.~(\ref{gammaexp}) shows, it will similarly
carry both signs, with the signs of all other quantities in 
Eq.~(\ref{gammaexp}) arguably being fixed. 
If $\Im(\gamma) >0$, then stability will be achieved only
by $\Delta m>0$, i.e. if $m> m_{\mathrm c}$. This is to say 
that a polygon, so formed, will be stable only if the number
of its vertices is above a threshold given by $m_{\mathrm c}$. 
The opposite argument applies if $\Im(\gamma) <0$, because in 
this case stability is ensured by $m<m_{\mathrm c}$, with there 
being an upper limit to the vertices of a stable polygon. 

Recent experiments by~\citet{tm15} have shown unstable polygons 
to undergo a rotational behaviour. We reproduce this feature 
theoretically with the help of $\Re(\gamma)$, in which the sign 
of $m_{\mathrm c}$ controls the clockwise or the anticlockwise 
direction of the rotation. The angular frequency of the rotation 
is $\Omega_{\mathrm{rot}} =\vert \Re(\gamma)\Delta m\vert$. 
In an unstable 
situation where $m< m_{\mathrm c}$, small values of $m$ yield
high values of $\vert \Delta m \vert$, and so 
$\Omega_{\mathrm{rot}} \propto \vert m_{\mathrm c}-m\vert$. 
This linear 
decay of the angular frequency with increasing number of 
vertices matches experimental results~\cite{tm15}. To find 
how $\Omega_{\mathrm{rot}}$ depends on the flow rate, $Q$, 
we first note from Eq.~(\ref{gammaexp}) that 
$\Omega_{\mathrm{rot}} \sim v_0(r_{\mathrm J})/r_{\mathrm J}$, 
in which 
$r_{\mathrm J} \sim Q^{5/8} \nu^{-3/8} g^{-1/8}$, a well-known
scaling result~\cite{bdp93} that is derived by equating
the dynamic time scale of the steady radial flow with the time 
scale of viscous dissipation~\cite{rb07}. The steady 
background quantities depend on viscosity, and at the jump, 
where $v_0(r_{\mathrm J})=\sqrt{gh_0(r_{\mathrm J})}$, the 
flow height, $h_0(r_{\mathrm J})$, is scaled by combining 
the aforementioned scaling of $r_{\mathrm J}$ with the 
integral solution of Eq.~(\ref{bdpcon}). This results in 
$h_0(r_{\mathrm J}) \sim Q^{1/4} \nu^{1/4} g^{-1/4}$~\cite{bdp93}, 
with which we get the scale, 
$\Omega_{\mathrm{rot}} \sim Q^{-1/2} \nu^{1/2}g^{1/2}$.  
Evidently, $\Omega_{\mathrm{rot}}$ decreases 
with increasing flow rate, $Q$, something that has been observed 
by~\citet{tm15} in their experiments. Our theoretical treatment 
is, therefore, well in accord with two experimental findings
of~\citet{tm15}, namely, the two ways for $\Omega_{\mathrm{rot}}$ 
to decay --- with increasing
number of polygon vertices and with increasing flow rate. 
Beyond these two established facts, we make a prediction, based 
on $\Omega_{\mathrm{rot}} \propto \nu^{1/2}$, that the angular 
velocity of the rotating polygons will increase with increasing 
kinematic viscosity. 

Our most crucial claim is that the breaking of the 
axial symmetry has a connection with the Hawking radiation 
in an acoustic white hole. If the right hand side of 
Eq.~(\ref{fpert2}) were to vanish, the left hand side will 
bring forth the symmetric metric of an analogue white 
hole~\cite{su02,sbr05,rb07}, a
point of view that is applicable to a steady radial outflow. 
The $\phi$-dependent terms on the right hand side of 
Eq.~(\ref{fpert2}) break the axial symmetry of the steady
radial flow, and at the
horizon, where $v_0^2=gh_0$, the flow becomes azimuthal. 
The analogue ``surface gravity" at the 
horizon~\cite{un81,un95,vis98} is given as 
${\mathrm g}_{\mathrm s} = (-1/2)\times 
[{\mathrm d}(v_0^2-gh_0)/{{\mathrm d}r}]$, which, taken 
together with Eq.~(\ref{mcrit1}), means 
$m_{\mathrm c}^2 \sim [{\mathrm g}_{\mathrm s}/v_0(r_{\mathrm J})] 
\Omega_{\mathrm{rot}}^{-1}$. Further, in terms of 
${\mathrm g}_{\mathrm s}$, the Hawking temperature, 
$T_{\mathrm H}$, can be set down as~\cite{vis98}  
$k_{\mathrm B}T_{\mathrm H} = \hbar \Omega_{\mathrm H}$, 
in which $\Omega_{\mathrm H} = 
{\mathrm g}_{\mathrm s}/[2\pi v_0(r_{\mathrm J})]$  
is a characteristic frequency of the thermal
Hawking radiation. 
Clearly, $\Omega_{\mathrm{rot}} \sim \Omega_{\mathrm H}$, 
whose startling implication is that the frequency with which 
the polygon rotates is a practicably measurable manifestation 
of the Hawking effect in an analogue white hole, a statement 
that is in conformity with a very recent observation of 
Hawking radiation emanating from an analogue black hole 
in an atomic Bose-Einstein condensate~\cite{jeffs16}. 
The corresponding
Hawking temperature specifies a thermal scale for any 
energy involved in the Hawking process~\cite{tj91}, and 
at high frequencies, $\tilde{\omega} \gg \Omega_{\mathrm H}$, 
the tunnelling amplitude of the Hawking radiation~\cite{pw2k} 
assumes the recognizable form of the Boltzmann factor, 
$\exp [-(\hbar \tilde{\omega})/(k_{\mathrm B}T_{\mathrm H})]$.   

\bibliography{rsbb2016}
\end{document}